\documentstyle[pra,aps,epsf]{revtex}
\draft

\begin{document}

\title{Breit interaction correction to the hyperfine
constant of an external s-electron in many-electron atom}
\author{O.P. Sushkov}

\address{School of Physics, University of New South Wales,\\
 Sydney 2052, Australia}

%\date{\today}
\maketitle

\begin{abstract}
Correction to the hyperfine constant $A$ of an external s-electron
in many-electron atom caused by the Breit interaction is calculated
analytically: $\delta A/A =0.68 Z\alpha^2$.
Physical mechanism for this correction is polarization of the
internal electronic shells (mainly $1s^2$ shell) by the magnetic
field of the external electron. 
This mechanism is similar to the polarization of vacuum
considered  by Karplus and Klein \cite{KK} long time ago.
The similarity is the reason why in both cases (Dirac sea polarization and
internal atomic shells polarization) the corrections have the same
dependence on the nuclear charge and fine structure constant.\\
In conclusion we also discuss $Z\alpha^2$ corrections to the parity violation
effects in atoms.
\end{abstract}


\section{introduction}
Atomic hyperfine structure is caused by the magnetic interaction of the
unpaired external electrons with the nuclear magnetic moment.
There are two types of relativistic corrections to this effect.
The first type is a single particle correction caused by the
relativistic effects in the wave equation of the external electron 
\cite{BS}. This correction is of the order of $(Z\alpha)^2$,
where $Z$ is the nuclear charge and $\alpha$ is the fine structure constant.
Solving the Dirac equation one can find this correction analytically
in all orders in $(Z\alpha)^2$ \cite{SFK}. An alternative way to find this 
correction is direct numerical solution of the Dirac equation.
Correction of the second type has a many-body origin: it is due to  
polarization of paired electrons by the Breit interaction of the
external electron.
There are two kinds of paired electrons in the problem:
a) Dirac sea, b) closed atomic shells.
The contribution related to the Dirac sea was calculated 
by  Karplus and Klein long time ago \cite{KK}: 
$\delta A/A= \alpha/2\pi-Z\alpha^2(5/2-\ln 2)$.
The $\alpha/2\pi$ part is due to usual anomalous magnetic moment and
$Z\alpha^2$ part comes from a combined effect of the Nuclear Coulomb
field and the Breit interaction. We would like to stress that 
as soon as $Z > 12$ the $Z\alpha^2$ part is bigger than $\alpha/2\pi$ .
Effect of the polarization of atomic shells by the Breit
interaction has been recently calculated numerically \cite{B,S,D,DD}.
These calculations were performed
for Cs ($Z=55$) because they were motivated by the interest to parity
nonconservation in this atom. Results of these calculations are
somewhat conflicting, but nevertheless they indicate
that the correction for an external s-electron is of the order of
%few tenths of a per cent.
$\sim \pm 0.4\%$. 
In spite of being rather small this correction is comparable
with the present accuracy of atomic many-body calculations and therefore
it must be taken into account.
The purpose of the present work is to calculate analytically the
correction induced by the Breit interaction.
This allows to elucidate the physical
origin of the effect and its dependence on the atomic parameters.
This also provides an important lesson for a similar correction to
the parity non-conserving amplitude which we discuss in the conclusion.

\section{Contribution of the direct electron-electron magnetic interaction}
In the present work we do not consider single particle relativistic
effects (Dirac equation), so we assume that in zero approximation
the atom is described by the Schroedinger equation with Coulomb 
electron-nucleus and electron-electron interaction.
For magnetic interaction throughout this work we use the Coulomb gauge.
Vector potential and magnetic field of the nucleus are of the form
\begin{eqnarray}
\label{An}
&&{\bf A}_N(r)={{{\bf \mu}_N \times {\bf r}}\over{r^3}},\\
&&{\bf {\cal H}}_N(r)={\bf \nabla \times A}_N \to
{{8\pi}\over{3}}{{\bf \mu}_N} \delta({\bf r}),\nonumber
\end{eqnarray}
where ${\bf \mu_N}$ is the magnetic moment of the nucleus.
We keep only the spherically symmetric part of the magnetic field ${\cal H}_N$
because in this work we consider only s-electrons.
Interaction of the magnetic moment $\mu_1$ of the external electron with
nuclear magnetic field gives the hyperfine structure
\begin{eqnarray}
\label{Hhfs}
&&H_{hfs}=\langle -\mu_1 \cdot {\cal H}_N\rangle = -C \ 
({\bf \mu_1 \cdot \mu_N})=A \ ({\bf s \cdot I}), \\
&&C={{8\pi}\over{3}} \ \psi^2_e(0),\nonumber\\
&&A={{\mu_1}\over{s}}{{\mu_N}\over{I}}C.\nonumber
\end{eqnarray}
Here $\psi_e(r)$ is the wave function of the external electron,
and $A$ is the hyperfine constant.

Vector potential of the external s-electron is of the form
\begin{equation}
\label{A}
{\bf A}_e(r)=\int{{\mu_1 \times ({\bf r-R})}\over{|{\bf r-R}|^3}}
\ \psi_e^2(R)d^3R
=-{\bf \mu_1}\times {\bf {\nabla}}_r
\int{{\psi_e^2(R)}\over{|{\bf r-R}|}} \ d^3R=
4\pi \ {{{\bf \mu_1} \times {\bf r}}\over{r^3}} \ \int_0^r \psi_e^2(R)\ R^2 dR.
\end{equation}
Hence the magnetic field is
\begin{equation}
\label{He}
{\bf {\cal H}}_e(r)={\bf \nabla \times A}_e \to
{{8\pi}\over{3}} \psi^2_e(r) \ \mu_1.
\end{equation}
Let us repeat once more that we keep only the spherically symmetric
part of the magnetic field.

The Hamiltonian of an internal electron in the magnetic field 
${\bf A} = {\bf A}_N+{\bf A}_e$, ${\cal H}={\cal H}_N+{\cal H}_e$
is given by
\begin{equation}
\label{Hm}
H={{({\bf p}-{{e}\over{c}}{\bf A})^2}\over{2m}}-\mu_2\cdot{\cal H}
+U(r),
\end{equation}
where $\mu_2$ is the magnetic moment of the internal electron. Certainly
$|\mu_1|=|\mu_2|={{|e|\hbar}\over{2mc}}$, however directions of
$\mu_1$ and $\mu_2$ are independent. Having eq. (\ref{Hm}) in mind
one can easily draw diagrams describing correction to the
hyperfine structure due to the electron-electron magnetic interaction.
This diagrams are shown in Fig.1

\begin{figure}[h]
\vspace{2pt}
\hspace{-35pt}
\epsfxsize=12cm
\centering\leavevmode\epsfbox{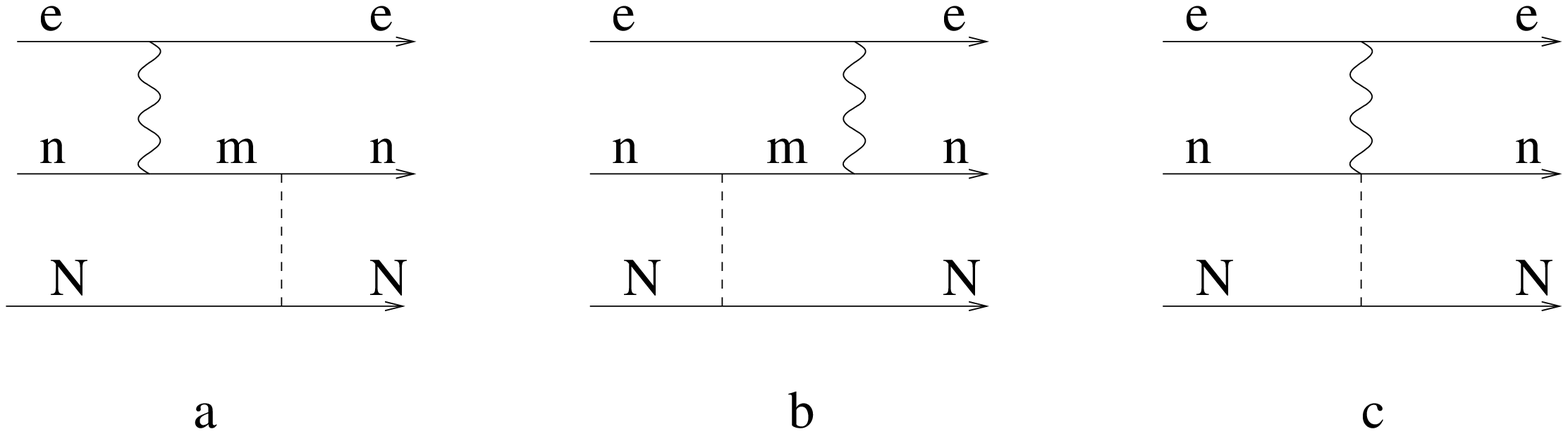}
\vspace{8pt}
\caption{\it {Diagrams describing the direct Breit interaction correction
to the hyperfine structure.
The diagrams a) and b) give the paramagnetic contribution, and
the diagram c) gives the diamagnetic contribution.
N denotes the nucleus, $e$ denotes the external electron, $n$ denotes
the internal electron, and finally $m$ denotes a virtual excitation of the
internal electron. The wavy line shows magnetic interaction with the
external electron, and the dashed line shows magnetic interaction with the
nucleus.}}
\label{Fig1}
\end{figure}

Two equal paramagnetic contributions are given by the diagrams shown in 
Fig.1a,b. Corresponding energy corrections are
\begin{equation}
\label{dEab}
\delta E_a =\delta E_b =\sum_{n \in filled}
\langle n|(-\mu_2\cdot{\cal H}_e|\delta \psi_{n}\rangle,
\end{equation}
where
\begin{equation}
\label{dpsin}
\delta \psi_{n} =\sum_{m\ne n} {{\langle m|(-\mu_2\cdot {\cal H}_N)|n\rangle}
\over{E_n-E_m}} |m\rangle.
\end{equation}
The diamagnetic contribution shown in Fig.1c is given by the $A^2$ term 
from the Hamiltonian (\ref{Hm}), hence
\begin{equation}
\label{dEc}
\delta E_c ={{e^2}\over{mc^2}}\langle n|{\bf A}_e \cdot {\bf A}_N|n\rangle.
\end{equation}

Before proceeding to the accurate calculation of $\delta E$ it is
instructive to estimate a magnitude of the correction. 
Let us look for example at the diamagnetic correction (\ref{dEc}).
According to eqs. (\ref{An}), (\ref{A}) the vector potentials are 
$A_N \sim \mu_N/r^2$ and $A_e \sim \mu_1 r \psi_e^2(0)$. 
Hence the correction (\ref{dEc}) is of the order of
$\delta E \sim e^2 \mu_1 \mu_N \psi_e^2(0)/(mc^2 r)$, where $r$
is the radius of the internal shell. Since all the interactions are
singular it is clear that the main contribution comes from the
K-shell, so $r \sim a_B/Z$ ($a_B$ is the Bohr radius). Together
with eq. (\ref{Hhfs}) this gives the following relative value of 
the Breit correction to the hyperfine constant $\delta C/C \sim Z\alpha^2$.
So we see that this effect has exactly the same dependence on the
atomic parameters as the Dirac sea polarization considered in the
paper \cite{KK}.

Now let us calculate the coefficient in the $Z\alpha^2$ correction.
We consider explicitly only 1s and 2s closed shells and we also need
to consider the external s-electron.
In atomic units the single particle energies of these states are: 
$E_1=-Z^2/2$, $E_2=-Z^2/8$, $E_e \approx 0$.
At small distances the nuclear Coulomb field is practically unscreened
and hence the wave functions are of the simple form \cite{LL}
\begin{eqnarray}
\label{wf}
&&\psi_1={1\over{\sqrt{\pi}}}e^{-\rho},\nonumber\\
&&\psi_2={1\over{\sqrt{8\pi}}}e^{-\rho/2}(1-\rho/2),\\
&&\psi_e=\sqrt{{3\over{16\pi\rho}}}J_1(\sqrt{8\rho}).\nonumber
\end{eqnarray}
Here $\rho=Zr$ and $J_1(x)$ is the Bessel function. The functions
$\psi_{1,2}$ are normalized in the usual way: $\int \psi_i^2 d^3 \rho =1$.
The wave function of the external electron is normalized by the
condition $\psi^2_e(0)=3/(8\pi)$. With this normalization the leading order
hyperfine constant (\ref{Hhfs}) is equal to unity, $C=1$, and therefore
this normalization is convenient for calculation of the
relative value of the Breit correction to the hyperfine constant.
Using eqs. (\ref{An}),(\ref{A}),(\ref{dEc}) and performing summation
over spins in the closed shells one finds the diamagnetic correction
\begin{equation}
\label{dEc1}
\delta E_c=Z\alpha^2 (\mu_1\cdot \mu_N)\sum_n\left({4\over{3}}
\int_0^{\infty}{{\psi_n^2(\rho)}\over{\rho^4}}d^3 \rho \int_0^{\rho}
\psi_e^2(\rho')d^3\rho'\right)=0.230 Z\alpha^2 (\mu_1\cdot \mu_N).
\end{equation}
The numerical coefficient was found by straightforward numerical integration.
Contributions of the inner shells drop down approximately as $1/n^3$, so
0.230=0.207+0.023, where the first contribution comes from the 1s-shell and the
second contribution comes from the 2s-shell.

To calculate the paramagnetic contributions (\ref{dEab}) we use
corrections $\delta\psi_n$  defined by eq. (\ref{dpsin}) and calculated
in the Appendix. Substitution of
(\ref{He}), (\ref{wf}), and (\ref{psin}) into formula (\ref{dEab})
and summation over electron spins in the closed shell
gives the following result
\begin{eqnarray}
\label{dEab1}
\delta E_a +\delta E_b &=&-{8\over{3}}Z\alpha^2 (\mu_1\cdot\mu_N)
\left(\int_0^{\infty}e^{-2\rho}J_1^2(\sqrt{8\rho})w_1(\rho)d\rho
+{1\over{8}}\int_0^{\infty}e^{-\rho}(1-\rho/2)J_1^2(\sqrt{8\rho})w_2(\rho)d\rho
\right)\\
&=&-{8\over{3}}Z\alpha^2 (\mu_1\cdot\mu_N) (0.219+0.021)=
-0.640Z\alpha^2 (\mu_1\cdot\mu_N),\nonumber
\end{eqnarray}
where we present explicitly the contributions of 1s- and 2s-shells.
The numerical coefficient is found by  numerical integration.
Similar to the diamagnetic term the contributions of the inner shells 
drop down approximately as $1/n^3$.

The leading order hyperfine structure is given by eq. (\ref{Hhfs})
with constant $C=1$ due to the accepted normalization.
According to eqs. (\ref{dEc1}) and (\ref{dEab1}) the total correction caused
by the direct magnetic interaction is 
$\delta E_a+\delta E_b +\delta E_c= -0.410Z\alpha^2(\mu_1\cdot\mu_N)$.
Comparing this with eq. (\ref{Hhfs}) one finds the relative value of the direct
correction:
\begin{equation}
\label{dcd}
{{\delta A^{(dir)}}\over{A}}={{\delta C^{(dir)}}\over{C}}=0.410Z\alpha^2.
\end{equation}

\section{Contribution of the exchange electron-electron magnetic interaction}
Exchange diagrams contributing to the correction are shown in Fig2.

\begin{figure}[h]
\vspace{2pt}
\hspace{-35pt}
\epsfxsize=12cm
\centering\leavevmode\epsfbox{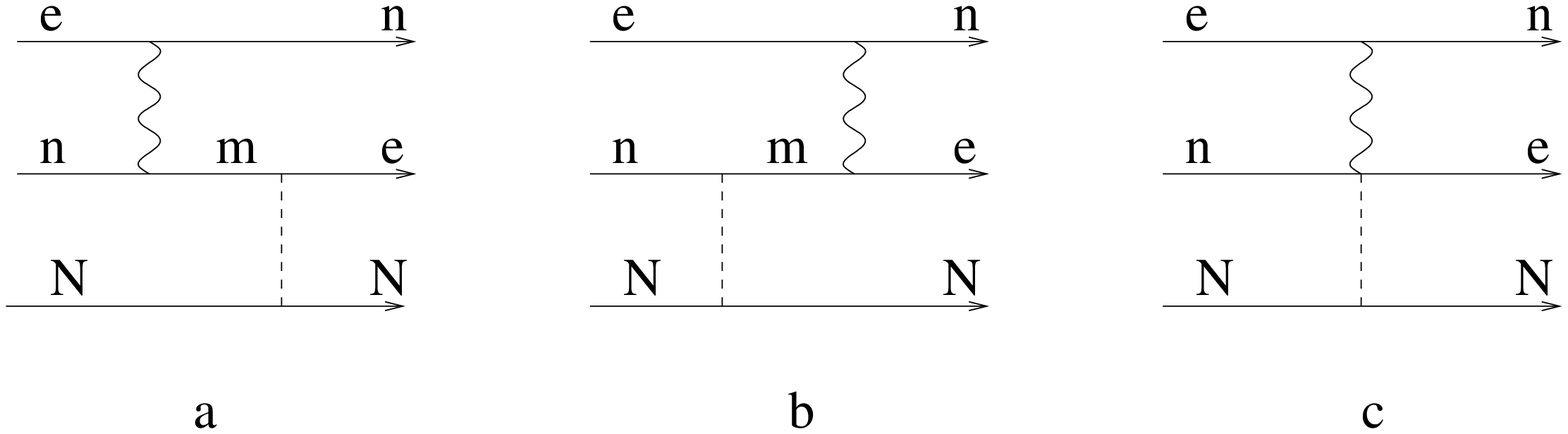}
\vspace{8pt}
\caption{\it {Diagrams describing the exchange Breit interaction correction
to the hyperfine structure.
The diagrams a) and b) give the paramagnetic contribution, and
the diagram c) gives the diamagnetic contribution.
N denotes the nucleus, $e$ denotes the external electron, $n$ denotes
the internal electron, and finally $m$ denotes a virtual excitation of the
internal electron. The wavy line shows magnetic interaction with the
external electron, and the dashed line shows magnetic interaction with the
nucleus.}}
\label{Fig2}
\end{figure}

The diagrams Fig.2a,b show the ``paramagnetic'' contributions, and
the diagram Fig.2c shows the ''diamagnetic'' contribution.
Note that the contributions of the diagrams Fig.2a,b must be doubled
because the opposite order of operators is also possible.
Let us begin with the ``dimagnetic'' term. Comparison of Fig.1c and
Fig.2c shows that the direct and the exchange contributions
are very similar and therefore the simplest way to derive the exchange 
term is just to make appropriate alterations in eq. (\ref{dEc1}) which gives
the direct contribution. The alterations are obvious: 1)opposite sign,
2)$\psi_n^2 \to \psi_e \psi_n$, 3)$\psi_e^2 \to \psi_e \psi_n$,
4)there is no summation over the intermediate spins, hence $4/3 \to 2/3$.
Thus the result is
\begin{equation}
\label{dEc1e}
\delta E^{(ex)}_c=-Z\alpha^2 (\mu_1\cdot \mu_N)\sum_n\left({2\over{3}}
\int_0^{\infty}{{\psi_n(\rho)\psi_e(\rho)}\over{\rho^4}}d^3 \rho 
\int_0^{\rho}\psi_n(\rho)\psi_e(\rho')d^3\rho'\right)=
-0.107 Z\alpha^2 (\mu_1\cdot \mu_N).
\end{equation}
The coefficient is found by  numerical integration:
0.107=0.096+0.011, where the first contribution comes from the 1s-shell and 
the second contribution comes from the 2s-shell.

The paramagnetic exchange contribution shown in Fig.2b is similar to the
direct ones given by Fig.1a,b. The only difference is in algebra of Pauli
matrixes and in additional sign (-). This consideration shows that the
paramagnetic exchange contribution is equal to half of that given by eq.
(\ref{dEab1}) 
\begin{eqnarray}
\label{dEb2}
\delta E_b^{(ex)}&=&-{4\over{3}}Z\alpha^2 (\mu_1\cdot\mu_N)
\left(\int_0^{\infty}e^{-2\rho}J_1^2(\sqrt{8\rho})w_1(\rho)d\rho
+{1\over{8}}\int_0^{\infty}e^{-\rho}(1-\rho/2)J_1^2(\sqrt{8\rho})w_2(\rho)d\rho
\right)\\
&=&-0.320Z\alpha^2 (\mu_1\cdot\mu_N).\nonumber
\end{eqnarray}
Note that the sign of the exchange contribution is the same as the sign of the
direct one (\ref{dEab1}).

The diagram shown in Fig.2a does not have analogous direct diagram because
it has the hyperfine interaction attached to the line of the external
electron. Nevertheless the calculation of this diagram is quite similar to 
the calculation described by eqs. (\ref{dEab}) and (\ref{dpsin}).
After substitution of $\delta\psi_e$ from (\ref{psin2}) and performing
summation over the polarizations inside the closed shell one finds the 
following expression for the diagram shown in Fig.2a
\begin{eqnarray}
\label{dEa2}
\delta E_a^{(ex)}&=&-8\pi Z\alpha^2 (\mu_1\cdot\mu_N)
\left(\int_0^{\infty}e^{-2\rho}J_1(\sqrt{8\rho})N_1(\sqrt{8\rho})\rho d\rho
+{1\over{8}}\int_0^{\infty}e^{-\rho}(1-\rho/2)^2
J_1(\sqrt{8\rho})N_1(\sqrt{8\rho})\rho d\rho
\right)\\
&=&0.156Z\alpha^2 (\mu_1\cdot\mu_N).\nonumber
\end{eqnarray}

The total exchange magnetic correction is
$\delta E^{(ex)}_a+\delta E^{(ex)}_b +\delta E^{(ex)}_c= 
-0.271Z\alpha^2(\mu_1\cdot\mu_N)$.
Comparing this with eq. (\ref{Hhfs}) one finds the relative value of the 
exchange correction:
\begin{equation}
\label{dcex}
{{\delta A^{(ex)}}\over{A}}={{\delta C^{(ex)}}\over{C}}=0.271Z\alpha^2.
\end{equation}

\section{Total Breit correction.\\
$Z\alpha^2$ correction due to electron-electron Coulomb interaction}

Adding the direct (\ref{dcd}) and the exchange (\ref{dcex}) contributions one 
finds the total Breit correction
\begin{equation}
\label{dctot}
{{\delta A_B}\over{A}}={{\delta C_B}\over{C}}=0.681Z\alpha^2.
\end{equation}
In the calculation we have not used the explicit form of the Breit
interaction, but nevertheless this is the correction generated by the
interaction which reads in the relativistic form and in the Coulomb gauge
as (see ref. \cite{BS})
\begin{equation}
\label{B}
H_B=-{1\over{2r}}\left( {\bf \alpha_1}\cdot {\bf \alpha_2}+
{{({\bf \alpha_1}\cdot {\bf r})({\bf \alpha_2}\cdot {\bf r})}\over{r^2}}
\right).
\end{equation}
Here ${\bf r=r_1-r_2}$ is distance between the electrons, and ${\bf \alpha}_i$
is the $\alpha$-matrix of the corresponding electron.
The  Breit interaction correction to the hyperfine structure of Cs 
was previously calculated numerically in the works\cite{B,S,D,DD},
but results of these calculations were somewhat conflicting.
Our result (\ref{dctot}) agrees  with that of the most recent
computation \cite{DD}. Note that eq. (\ref{dctot}) gives the leading
in $Z$ part of the Breit correction. There are other parts, say the
correction to the energy of the external electron which directly
influence the hyperfine constant. However the other parts contain
lower powers of $Z$.

The correction (\ref{dctot}) does not include all $Z\alpha^2$ terms.
To realize what is left let us look at the electron-electron
interaction Hamiltonian in $(v/c)^2$ approximation \cite{BLP}
\begin{eqnarray}
\label{BB}
H_B=&&\alpha^2\left\{-\pi \delta({\bf r})-{1\over{2r}}
\left(\delta_{\alpha \beta}+{{r_{\alpha} r_{\beta}}\over{r^2}}\right)
p_{1\alpha}p_{2\beta}
+{1\over{2r^3}}\left(-{\bf (s_1+2s_2)}\cdot[{\bf r}\times{\bf p_1}]
+{\bf (s_2+2s_1)}[{\bf r}\times{\bf p_2}]\right)\right.\\
&& \ \ \ \ \ \ \ +\left.{{\bf s_1\cdot s_2}\over{r^3}}
-{{3(\bf s_1\cdot r)(s_2\cdot r)}\over{r^5}}
-{{8\pi}\over{3}}{\bf s_1\cdot s_2}\delta({\bf r})\right\}.\nonumber
\end{eqnarray}
Here ${\bf p_i, s_i}$ denote the momentum and the spin of the electron.
All the terms containing momenta vanish for s-electrons, the two
last terms are already taken into account by
the calculation performed above, however the first term has not
been considered yet. The matter is that in spite of being a 
$(v/c)^2$-correction it has a nonmagnetic origin. It comes
from the $(v/c)$-expansion of the electron-electron Coulomb interaction
${1\over{r}}(u_1^{\dag}u_1)(u_2^{\dag}u_2)$, where $u_i$ is the Dirac
spinor of the corresponding electron.
This is why this term is accounted automatically in the Dirac-Hartree-Fock
calculations \cite{DFKS,B,S,D,DD}. Nevertheless if one wants to separate
the total $Z\alpha^2$ correction analytically then the first term must be
also considered explicitly.
Since this term is spin independent,
it can contribute only via exchange diagrams.
A straightforward calculation very similar to that performed above gives
the following result for the Coulomb correction.
\begin{eqnarray}
\label{dECoul}
\delta E_{Coulomb}&=&2Z\alpha^2 (\mu_1\cdot\mu_N)
\left(\int_0^{\infty}e^{-2\rho}J_1^2(\sqrt{8\rho})w_1(\rho)d\rho
+{1\over{8}}\int_0^{\infty}e^{-\rho}(1-\rho/2)J_1^2(\sqrt{8\rho})w_2(\rho)d\rho
\right)\nonumber\\
&-&4\pi Z\alpha^2 (\mu_1\cdot\mu_N)
\left(\int_0^{\infty}e^{-2\rho}J_1(\sqrt{8\rho})N_1(\sqrt{8\rho})\rho d\rho
+{1\over{8}}\int_0^{\infty}e^{-\rho}(1-\rho/2)^2
J_1(\sqrt{8\rho})N_1(\sqrt{8\rho})\rho d\rho
\right)\\
&=&0.558Z\alpha^2 (\mu_1\cdot\mu_N).\nonumber
\end{eqnarray}
Comparing this with eq. (\ref{Hhfs}) one finds the relative value of the 
Coulomb correction:
\begin{equation}
\label{dCoul}
{{\delta A_{Coulomb}}\over{A}}={{\delta C_{Coulomb}}\over{C}}=-0.558Z\alpha^2.
\end{equation}
Combination of the magnetic (\ref{dctot}) and of the Coulomb (\ref{dCoul})
corrections give the total $Z\alpha^2$ correction to the hyperfine
constant due to the polarization of the closed electronic shells.
\begin{equation}
\label{ddt}
{{\delta A}\over{A}}=0.123Z\alpha^2.
\end{equation}

\section{Breit interaction correction to the parity nonconservation effect}
The Breit interaction correction exists for both nuclear spin independent
(weak charge) and nuclear spin dependent (anapole moment) weak 
interactions. The correction to the spin independent effect is the 
most interesting one because of the high precision in both atomic theory 
\cite{DFS} and experiment \cite{W,BW}.
Recent computations \cite{D,DD} show that the atomic Breit correction
to the nuclear spin-independent parity nonconservation (PNC) effect  in Cs
is about 0.6\%. This is enough to influence interpretation of
the experimental data, see Refs. \cite{BW,D}.
The Breit correction to the PNC effect can be calculated analytically
similar to the hyperfine structure correction. However this calculation is
out of the scope of the present work. In the present paper I just want
a) to estimate this correction parametrically, b) to comment on the importance
of the Dirac sea polarization.
Let us first estimate the correction. It is given by the diagrams similar
to that shown in Fig.2a,b. The only difference is that the electron-nucleus
magnetic interaction must be replaces by the electron-nucleus weak
interaction \cite{Kh}
\begin{equation}
\label{Hw}
H_W={{G}\over{2\sqrt{2}m}}Q_W [{\bf s \cdot p}\delta({\bf r})
+\delta({\bf r}){\bf s \cdot p}],
\end{equation}
where $G$ is the Fermi constant, and $Q_W$ is the weak charge.
The relative value of the Breit correction is
\begin{equation}
\label{dHw}
\delta\langle H_W\rangle \left/\right.\langle H_W\rangle \sim
{{\alpha^2/r^3}\over{\Delta E}} \sim
{{\alpha^2 Z^3}\over{Z^2}} \sim Z \alpha^2.
\end{equation}
Here $\alpha^2/r^3$ is the magnetic interaction between external and 
internal electrons, and $\Delta E$ is the excitation energy of the
virtual state of the internal electron. For the estimation we take
K-electrons, therefore $r \sim 1/Z$, and $\Delta E \sim Z^2$.
Thus the Breit correction to the PNC effect has exactly the same
dependence on atomic parameters as the Breit correction to the
hyperfine structure. It is known that there is a large relativistic
enhancement factor for PNC effect \cite{Kh}, therefore one might think
that the nonrelativistic expansion used in the estimate (\ref{dHw})
has very poor accuracy. However it is not the case, the matter is that
the large relativistic factor appears from the distances 
$r \sim nuclear \ \ size \ll a_B/Z$, therefore this factor is more or less the 
same for the external electron and for the K-electron. So it is canceled out 
in the ratio (\ref{dHw}).
By the way a similar argument explains a relatively high accuracy
of the nonrelativistic expansion for the hyperfine structure correction
(\ref{dctot}).

I would like also to comment on the importance of the Dirac sea polarization.
The effect calculated in refs.\cite{D,DD} and estimated in 
eq. (\ref{dHw}) is caused by the polarization of the closed atomic shells.
This is analogous to the contribution (\ref{dctot}) for the hyperfine
structure. However one has to remember that for the hyperfine correction
there is also the contribution of the Dirac sea polarization \cite{KK},
$\alpha/2\pi-Z\alpha^2(5/2-\ln 2)\approx \alpha/2\pi-1.81 Z\alpha^2 $,
which is bigger  and has the opposite sign.
Only account of both effects together (atomic shells + Dirac sea) has
the physical meaning.
The radiative correction to the nuclear weak charge is known in 
the single loop approximation, see Ref. \cite{MS}. This includes
$\alpha/\pi$ and $\alpha/\pi\ln(M_Z/\mu)$ terms ($M_Z$ is the
Z-boson mass and $\mu$ is some infrared cutoff).
However $Z\alpha^2$ radiative correction 
to the weak charge has not been considered yet and it is quite
possible that this correction is larger than the $\alpha/\pi$ contribution
(at least this is so for the hyperfine constant).
 To calculate $Z\alpha^2$ radiative correction one needs to go to
the two loop approximation, or to work in the single loop approximation
but with the Green's functions in the external nuclear Coulomb field.
Thus: 1) account of the Breit correction calculated numerically in
Refs. \cite{D,DD} and estimated in eq. (\ref{dHw})
without account of the Dirac sea contribution is not sufficient,
2) account of the Dirac sea polarization can influence
agreement between theory and experiment.

\section{conclusion}
Correction to the hyperfine constant $A$ of an external s-electron
in many-electron atom caused by the polarization of inner atomic shells
by the electron-electron Breit interaction
is calculated analytically: $\delta A/A =0.68 Z\alpha^2$.
This correction has the same origin as the Dirac sea polarization effect
$\delta A/A= \alpha/2\pi-Z\alpha^2(5/2-\ln 2)$ calculated
by  Karplus and Klein long time ago\cite{KK}.

It has been shown that the parametric estimate for the Breit
correction to the parity nonconservation effects is also $Z\alpha^2$.
We stress that to take this correction into account one needs to
consider both polarization of the inner atomic shells and polarization
of the Dirac sea.

\acknowledgments
I am grateful to V. A. Dzuba for very important stimulating discussions.
I am also grateful to V. A. Dzuba and W. R. Johnson for communicating me
results of their calculations \cite{DD} prior to the publication.

\appendix
\section{corrections to the electronic wave functions
due the hyperfine interaction}
Correction $\delta\psi$ to the single particle wave function is given by 
eq. (\ref{dpsin}). Using this formula one can easily prove that
$\delta\psi$ satisfies the following equation
\begin{equation}
\label{St}
(H_0-E_n)\delta\psi_n=-\sum_{m\ne n}|m\rangle\langle m|V|n\rangle
=-V|n\rangle+\langle n|V|n\rangle|n\rangle.
\end{equation}
Here
\begin{equation}
\label{V}
V=-\mu \cdot {\cal H}_N =-{{8\pi}\over{3}} (\mu \cdot \mu_N) \delta({\bf r})
\end{equation}
is the perturbation and
\begin{equation}
\label{H0}
H_0={{p^2}\over{2m}}-{{Ze^2}\over{r}}\to -{1\over{2}}\Delta -{Z\over r}
\end{equation}
is the Hamiltonian of the Coulomb problem.
The screening of the Coulomb field is neglected because we consider only  
small distances, $r \sim 1/Z$.
The eq. (\ref{St}) has an infinite set of solutions. To find the correct one
we have to remember that there is the additional condition of orthogonality
\begin{equation}
\label{ort}
\langle \delta \psi_n|\psi_n\rangle=0,
\end{equation}
which follows from eq. (\ref{dpsin}).
Having the perturbation (\ref{V}) it is convenient to represent $\delta\psi_n$
in the form 
\begin{equation}
\label{psif}
\delta\psi_n={{4}\over{3}}Z^{5/2}(\mu \cdot \mu_N)\psi_n(0)f_n(\rho),
\end{equation}
where $\rho =Zr$.
Substitution of (\ref{psif}) into (\ref{St}) shows that the functions $f(\rho)$
obey the following equations
\begin{eqnarray}
\label{ff}
&&1s: \ \ \left(\Delta_{\rho}+{{2}\over{\rho}}-1 \right)f_1(\rho)
=-4\pi \delta(\rho)+4e^{-\rho}, \\ 
&&2s: \ \ \left(\Delta_{\rho}+{{2}\over{\rho}}-{1\over{4}} \right)f_2(\rho)
=-4\pi \delta(\rho)+{1\over{2}}e^{-\rho/2}(1-\rho/2).
\nonumber
\end{eqnarray}
To satisfy the boundary conditions at $\rho=0$ and at$\rho=\infty$
it is convenient to use another substitution 
\begin{eqnarray}
\label{subs}
&&f_1={1\over{\rho}}e^{-\rho}w_1(\rho),\\
&&f_2={1\over{\rho}}e^{-\rho/2}w_2(\rho),\nonumber
\end{eqnarray}
where $w_i(0)=1$ and $w_i(\rho)$ grows at large $\rho$ not faster than a 
polynomial. Straightforward solution of eqs. (\ref{ff}) together
with the orthogonality condition (\ref{ort}) gives
\begin{eqnarray}
\label{ws}
&&w_1=1-2\rho[\ln(2\rho)-5/2+c]-2\rho^2,\\
&&w_2=1-2\rho[(1-\rho/2)\ln \rho-3/4+c]-(13/4-c)2\rho^2+\rho^3/4,\nonumber
\end{eqnarray}
where $c=0.577215$ is the Euler constant.
Altogether eqs. (\ref{psif}), (\ref{subs}), and (\ref{ws}) give
\begin{eqnarray}
\label{psin}
&&\delta\psi_1={4\over{3\sqrt{\pi}}}Z^{5/2}(\mu \cdot \mu_N)
{1\over{\rho}}e^{-\rho}w_1(\rho),\\
&&\delta\psi_2={4\over{3\sqrt{8\pi}}}Z^{5/2}(\mu \cdot \mu_N)
{1\over{\rho}}e^{-\rho/2}w_2(\rho).\nonumber
\end{eqnarray}

We also need to know the hyperfine correction to the wave function
of the external electron. Basically it is also given by eq. (\ref{St}),
but there is a special point concerning normalization. 
We use an artificial normalization condition $\psi^2_e(0)=3/(8\pi)$, see 
eq. (\ref{ws}). On the other hand the correct normalization is
$\int \psi_e^2(r)d^3 r=1$ and hence $\psi_e(0)\propto E^{3/2}$, where
$E$ is energy of the electron. There are two terms in the right hand side 
of eq. (\ref{St}), the first term is proportional to $E^{3/2}$ and the second
one is proportional to $E^{9/2}$. For an external electron $E \to 0$
and hence the second term must be neglected. After that the equation
(\ref{St}) is getting linear in $\psi$ and we can return to the
normalization $\psi^2_e(0)=3/(8\pi)$.
Similar to (\ref{psif}) it is convenient to use the substitution
\begin{equation}
\label{psif1}
\delta\psi_e(r)=\sqrt{{{2}\over{3\pi}}}Z(\mu \cdot \mu_N)f_e(\rho),
\end{equation}
where $f_e$ satisfies the equation
\begin{equation}
\label{ff1}
\left(\Delta_{\rho}+{{2}\over{\rho}} \right)f_e(\rho)
=-4\pi \delta(\rho).
\end{equation}
Note that due to the different normalization of $\psi_e$ the coefficient 
in the right hand side of eq. (\ref{psif1}), including power of $Z$
is different from that in eq. (\ref{psif}).
Solution of eq. (\ref{ff1}) is
\begin{equation}
\label{NN}
f_e=\pi\sqrt{2\over{\rho}}N_1(\sqrt{8\rho}),
\end{equation}
where $N_1(x)$ is the singular Bessel function. At small $x$ this function
behaves as $N_1 \approx -2/\pi x$, therefore $f_e(\rho)$ has correct behavior
at small $\rho$: $f_e(\rho) \approx 1/\rho$.
Together with (\ref{psif1}) this gives
\begin{equation}
\label{psin2}
\delta\psi_e=-\sqrt{{{4\pi}\over{3}}}Z(\mu \cdot \mu_N)
{{N_1(\sqrt{8\rho})}\over{\sqrt{\rho}}}.
\end{equation}
To make sure that this is the correct solution one has to prove 
validity of the orthogonality condition (\ref{ort}). With eqs. (\ref{wf})
and (\ref{psin2}) one finds that the overlapping is of the form
\begin{equation}
\label{ort1}
\langle \delta \psi_n|\psi_n\rangle \propto \int_0^{\infty}
J_1(\sqrt{8\rho}) N_1(\sqrt{8\rho}) \rho d\rho \propto
\int_0^{\infty} J_1(x) N_1(x) x^3 dx.
\end{equation}
This integral is not well defined at $\infty$. The origin for this is
clear: we are working with zero energy state. 
To correct the situation, one has to introduce an exponential factor
$e^{-\beta x^2}$ and then to consider the limit $\beta \to 0$.
\begin{equation}
\label{ort2}
\langle \delta \psi_n|\psi_n\rangle \propto \lim_{\beta \to 0}
\int_0^{\infty}e^{-\beta x^2} J_1(x) N_1(x) x^3 dx
\propto \lim_{\beta \to 0}{1\over{\beta^5}}e^{-1/2\beta} W_{3/2,3/2}(1/\beta)
\propto \lim_{\beta \to 0}{1\over{\beta^{13/2}}}e^{-1/\beta} =0.
\end{equation}
For the evaluation of the integral we have used ref. \cite{GR}.
Formula (\ref{ort2}) completes the prove of the orthogonality.

\end{document}